# Chiral-reversing vortex radiation from a single emitter by eigenstates phase locking

Xing-Yuan Wang[1,2]‡, Hua-Zhou Chen[1]‡, Suo Wang[1], Li Ge[3]*, Shuang Zhang[4], Ren-Min Ma[1,2]*

[1] State Key Lab for Mesoscopic Physics and School of Physics, Peking University, Beijing 100871, China

[2] Collaborative Innovation Center of Quantum Matter, Beijing 100871, China

[3] Department of Engineering Science and Physics, College of Staten Island, CUNY, Staten Island, NY 10314, USA and the Graduate Center, CUNY, New York, NY 10016, USA

[4] School of Physics and Astronomy, University of Birmingham, Birmingham, B15 2TT, UK

‡These authors contributed equally to this work.

*Correspondence should be addressed to renminma@pku.edu.cn, li.ge@csi.cuny.edu

**ABSTRACT:**

The radiation of an emitter does not depend only on its intrinsic properties but also on the surrounding photonic environment, the notion of which is essential in the developments of lasers, quantum optics and other light-matter interaction related fields. However, in conventional wisdom, an emitter radiates into photonic eigenstates in the weak coupling regime and does not alter the property of the latter. Here, we report a counterintuitive phenomenon where the radiation field of a dipole in a parity-time symmetric ring resonator displays the opposite handedness to the eigenstates of the system. This chiral-reversing radiation takes place at an exception point of the underlying non-Hermitian system, where the singularity at the exceptional point forces a phase locking of the coalesced eigenstates when interacting with the dipole emitter. Such an intriguing phenomenon has been employed to construct vortex radiation with controllable topological charge from a single quantum dot embedded plasminic nanocavity with Purcell enhancement factor up to 1000. Our scheme enriches the interesting physics of an exception point in the quantum region and may open a new paradigm for chiral quantum optics and vortex lasers at nanoscale.

**INTRODUCTION**

In quantum information science, one of the prime tasks is to generate single photon states on demand from a single quantum emitter, such as an atom, a quantum dot or a nitrogen-vacancy center in diamond [1-7]. Cavity quantum electrodynamics (QED), which studies the strong interaction between a quantum emitter and radiation modes, has played a central role in this pursuit and developing practical sources of quantum states of light [1, 8]. For instance, a single quantum dot emitter coupled to a micropillar cavity has been employed in the recent demonstration of quantum boson-sampling machines with superior performance [9]. Another platform for solid-state cavity QED, namely plasmonic waveguides and cavities, have recently attracted growing interest in modifying radiation efficiency and directivity of single quantum emitters, where plasmonic effect with strong field localization enhances light-matter interaction significantly. [2-3, 10-14].

In the meanwhile, phase singularities or optical vortices have also received an ever increasing amount of attention from the optics community [15-28]. Most noticeably, devices that emit individual photons carrying orbital angular momentum (OAM) provide an exciting platform for using OAM in quantum information science, as they allow additional encoding on the single photon level [29-32]. Moreover, a multistate OAM system can be combined with spin angular momentum (SAM) or other degrees of freedom to form hyper entanglement or hybrid entanglement [33-37], which can significantly improve quantum computation, quantum communication, and quantum cryptography. As reported recently [38], a single photon encoded with both SAM and OAM has been utilized for quantum teleportation of composite states.

Notwithstanding the fast development of cavity QED in preparing single photon states, modulating the radiation pattern of a single emitter into a vortex beam with controllable topological charge remains a formidable task. While one can introduce chirality to the scattering light field of a nanoparticle or a nanoslit using circularly polarized light illumination in the classical regime, it is much more sophisticated to control the chirality of the radiation field from a single emitter in the quantum regime, where the Zeeman effect has to be introduced in the system, for instance [39-45]. A promising approach that has been demonstrated in the emergent chiral quantum optics [7] employs spin-momentum locking, i.e., placing a circularly polarized emitter in the vicinity of matching optical waveguides or cavities.

An alternative approach to introduce chiral light-matter interaction puts more emphasis on the photonic environment, which allows only unidirectional wave propagation. A novel class of chiral photonic structures are introduced using parity-time (PT) symmetry [46, 47] and its resultant non-Hermitian properties [48, 49]. PT symmetry requires an effectively balanced arrangement of optical gain and loss [50-61], and unidirectional reflectionless transmissions in a straight waveguide [62, 63] have been shown to be the result of a generalized flux conservation relation [64]. When wrapped into a ring, which would have two traveling-wave modes with opposite OAM in the absence of PT modulation, a single coalesced OAM mode emerges as the result of an exceptional point of the system Hamiltonian [65-66]. Such an optical exceptional point has been employed to construct chiral optical devices, including single mode lasers and vortex lasers [67-69].

Here, for the first time, we reveal a surprising phenomenon of chiral-reversing dipole radiation in a PT ring cavity at its exceptional point. Naively one would have expected the dipole radiation to follow the single OAM mode mentioned above. However, the singularity at the exceptional point prevents this conventional scenario from happening. Instead, we find that the chirality of the radiation pattern can be completely reversed from that of the OAM mode. Using a non-Hermitian perturbation theory, we show that the underlying mechanism can be understood as the phase locking between two nearly identical OAM modes: a $\pi$ phase shift is introduced by the exceptional point between their equal amplitude, which completely prohibits the emitter from radiating into their dominant handedness. Such an intriguing phenomenon is employed in the designing of a single quantum dot embedded chrial plasmonic nanocavity, which emanates vortex radiation with a controllable topological charge and a Purcell enhancement factor up to 1000. The phenomena of chiral-reversing vortex radiation by a single quantum dot enriches the interesting physics of an exception point in the quantum regime, and our scheme may open a new paradigm for chiral quantum optics and vortex lasers at nanoscale.

**RESULTS**

**Chiral-reversing dipole radiation induced by eigenstates phase locking**

The dipole radiation field can be directly calculated using the Green's function, which in turn can be expressed as a summation of all the eigenmodes in the system [Supplementary Note 1]:

$$G(\phi, \phi'; k) = \sum_m \frac{\psi_m(\phi)\psi_m(\phi')}{(k^2 - \tilde{k}_m^2)(m,m)} \equiv \sum_m \alpha_m \psi_m(\phi). \tag{1}$$

Here for simplicity we have treated the ring as one dimensional, and $\phi, \phi'$ are the angular position of the radiation field and emitter, respectively. $m$ is the mode index and $(m, n) = \int_0^{2\pi} n^2(\phi)\phi_m(\phi)\phi_n(\phi)d\phi$ is the non-Hermitian inner product, satisfying the biorthogonal relation $(m, n) = 0$. $n(\phi)$ is the refractive index of the ring, and $k$ is the wave vector of the dipole radiation in free space.

One important quantity in our analysis is $\alpha_m$, the amplitude of the radiation field in mode $m$. In conventional wisdom, $\alpha_m$ is large when the radiation frequency is at resonance with the corresponding cavity mode. For example, eigenstates appear in degenerate pairs in a uniform ring cavity, with clockwise (left handed) traveling wave in one mode and counterclockwise (right handed) traveling wave in the other. When a dipole emitter of the same frequency is introduced to the cavity, it excites both of the two on-resonance eigenstates and forms a standing wave radiation field inside the cavity (see the schematic in Fig. 1 and Supplementary Note 1).

When a weak PT-symmetric modulation is applied to the cavity, e.g., $n(\phi) = n_0 +$

$(\delta n_R \cos 2l\phi + i\delta n_I \sin 2l\phi)$ $(|\delta n_{R,I}| \ll n_0)$, the two originally degenerated eigenmodes with angular momentum $l$ become coupled and are given by [Supplementary Note 1]:

$$\psi_\pm(\phi) = \sqrt{1+\beta}e^{il\phi} \mp \sqrt{1-\beta}e^{-il\phi}. \tag{2}$$

An exceptional point occurs at $\beta \equiv \delta n_I/\delta n_R = 1$, where $\psi_\pm(\phi)$ coalesce to a single CCW eigenmode (Fig. 1b). Following the conventional wisdom mentioned above, one would expect naively that at resonance the dipole will radiate strongly into this eigenmode and hence the radiation field will inherit its chirality.

Remarkably, we find that the dipole emitter can radiate all of its power in the CW direction instead, i.e., exhibiting a chiral-reversing dipole radiation (Fig. 1c). To resolve this apparent dilemma, we first note that the eigenstates themselves do no form a complete basis at an exceptional point [70] and hence the expansion in Eq. (1) fails. To avoid this difficulty, we choose to shift the system slightly away from the exceptional point by letting $\beta = 1 - \varsigma^2$ ($0 < \varsigma \ll 1$), which lifts the degeneracy of $\psi_\pm(x)$. As Eq. (2) shows, these two nearly degenerate modes now both acquire a weak CW component (Fig. 1b), and their dominant CCW components must cancel each other in order to achieve a chiral-reversing dipole radiation. Extraordinarily, this property is forced by the diverging factors $\alpha_\pm$ in Eq. (1) (Supplementary Note 1): $\alpha_+ \approx -\alpha_- \propto \varsigma^{-1}$, which have the same (diverging) amplitude and are $\pi$ out of phase.

For a direct comparison, we have plotted dynamical three dimensional radiation fields of a dipole emitter inside a normal ring cavity, of the coalesced eigenstates of a ring cavity operating close to an exceptional point, and of a dipole emitter inside a ring cavity operating close to an exceptional point (Supplementary video 1).

This phase locking phenomenon is the main mechanism that leads to the chiral-reversing dipole radiation, and it is an intrinsic property of the Green's function at an exceptional point: any finite deviation of this phase difference from $\pi$ will result in an infinite radiation power due to the diverging amplitudes $\alpha_\pm$. In the meanwhile, the minute CW components in these two modes (proportional to $\pm\sqrt{1-\beta} \approx \pm\varsigma$) are enhanced by this singularity and they are *in phase*, which then determines the chirality of the radiation field. This phenomena is in stark contrast with previously reported chiral optical devices at an exceptional point, such as unidirectional reflectionless waveguide, single mode lasers and vortex lasers, where only the properties of the coalesced eigenstate are considered [63, 67-69].

We should note that a more rigorous calculation will produce a small CCW component in the radiation field, which is due to the higher order terms in $\varsigma\alpha_\pm$. Nevertheless, it can be easily eliminated as we show below, using an equivalent analysis but from a different perspective, i.e., by considering the interference between the instantaneous dipole radiation fields and the circulating cavity fields.

**Single emitter embedded chiral plasmonic nanocavity**

Fig. 2a illustrates the design of a chiral plasmonic nanocavity (CPN) operating at an exception point, which is a ring resonator with a metal-insulator-metal coaxial geometry. The bottom of the insulator ring is encapsulated by silver and with a patterned layer to introduce parity-time (PT) symmetric refractive index modulation (Fig. 2b). A single quantum dot is embedded inside the insulator region, and it is at resonance with a pair of whispering-gallery modes (WGMs). The electric field in the modulated regime can be written as: $E(\rho,\varphi,z) = U(\rho,z)[a_{\text{CW}}(t)e^{im\varphi} + a_{\text{CCW}}(t)e^{im\varphi}]$. To describe our single emitter embedded CPN, we formulate the coupled mode equations below by including (i) the coupling between the single emitter and the two degenerated counter-propagating WGMs; and (ii) the coupling between these two WGMs (electric field):

$$\frac{d}{dt}a_{\text{CW}} = -i\omega a_{\text{CW}} - \gamma_{tot}a_{\text{CW}} + \chi_{ab}a_{\text{CCW}} + \epsilon s,$$

(3)

$$\frac{d}{dt}a_{CCW} = -i\omega a_{CCW} - \gamma_{tot}a_{CCW} + \chi_{ba}a_{CW} + \epsilon s.$$

Here $a_{CW}$ ($a_{CCW}$) is the amplitude of the CW (CCW) mode, $\omega$ is the traveling WGM resonance frequency, $\gamma_{tot}$ is the total loss rate of the cavity, and $\chi_{ab}$ ($\chi_{ba}$) is the coupling coefficient from the CCW (CW) mode to the CW (CCW) mode. The dipole appears as a driving term in Eq. (3), represented by the instantaneous radiation amplitude $s$ and the coupling coefficient $\epsilon$ to the cavity fields (Supplementary Note 2 and Supplementary Fig. 2).

Under the condition of $s=0$ (no source inside the cavity), we can solve the (complex) resonant frequencies of the chiral plasmonic nanocavity from Eq. (3) as

$$\Omega_{\pm} = \omega - i\gamma_{tot} \pm i\sqrt{\chi_{ab}\chi_{ba}}, \tag{4}$$

and the two corresponding eigenstates are given by

$$\begin{pmatrix} a_{cw} \\ a_{ccw} \end{pmatrix} = \begin{pmatrix} \pm\sqrt{\chi_{ab}} \\ \sqrt{\chi_{ba}} \end{pmatrix}. \tag{5}$$

This result conveys the physics in Eq. (2) more transparently, i.e., the exceptional point is reached when one of the two coupling coefficients ($\chi_{ab}$, $\chi_{ba}$) equals zero, where the two eigenmodes given by Eq. 5 coalesce into a single OAM mode.

Instead of implementing the sinusoid PT-symmetric refractive index modulation mentioned in the previous section, we approximate it by a square waveform (Fig. 2b) to simplify the fabrication process. The corresponding coupling coefficients are proportional to the Fourier transform coefficients of $\delta n(\varphi)$ with angular momentum $\pm 2l$, which can be calculated as

$$\chi_{ab,ba} = \kappa(\delta n_I \mp \delta n_R)e^{i2l\varphi_0}. \tag{6}$$

Here $\kappa$ is a dimensionless constant and $e^{i2l\varphi_0}$ is a phase factor determined by the position of the dipole emitter (See Supplementary Note 2). Clearly, the backscattering is unidirectional when $\delta n_I$ equals $\delta n_R$ (i.e., $\beta = 1$), with $\chi_{ab} = 0$ but $\chi_{ba} \neq 0$. Consequently, the eigenstates coalesces to one CCW (right handed) mode.

Our emitter embedded CPN cavity exhibits the chiral-reversing dipole radiation at exceptional point mentioned in the previous section: The amplitude ratio of CCW and CW waves in the radiation field can be calculated using Eq. 3, and it is given by

$$a_{CCW}/a_{CW} = 1 + \frac{\chi_{ba}}{\gamma_{tot}} \quad (7)$$

in the steady state when the dipole is at resonance and the CPN is at the exceptional point. This ratio vanishes at $\chi_{ba} = -\gamma_{tot}$ (Fig. 2d), indicating the power of the dipole emitter all couples to the opposite direction of the CCW mode, which is the only eigenmode of the system at the exceptional point (See Supplementary Fig. 3 and Fig. 4).

Notably, this chiral-reversing vortex radiation phenomena not only requires the system operating at exceptional point ($\chi_{ab} = 0$), but also the additional condition of $\chi_{ba} = -\gamma_{tot}$, which has not been revealed before to the best of our knowledge. This additional condition can be understood by using the perturbation theory presented in the previous section and including the next order corrections (Supplementary Note 1). Here however, this additional requirement *and* the chiral-reversing dipole radiation are the manifestations of the completely destructive interference between the directly radiated and the backscattered CCW waves. Fig. 2e verifies this observation by performing a temporal evolution of $a_{cw,ccw}$ as the steady state is reached.

**Chiral-reversing vortex radiation from a single emitter**

The phenomenon of chiral-reversing dipole radiation not only presents the interesting physics at an exception point; it can also be utilized to manipulate a single quantum dot radiation in an unprecedented manner. Below we show the construction of vortex radiation with controllable topological charge from a single emitter.

The mechanism underlying the free space vortex beam generation from the CPN is the similarity between the cavity field and the free space vortex beams, given by WGMs and Bessel-Gauss beams respectively. Both of them consist of Bessel functions in the radial direction and a phase factor of $e^{i(l\varphi - k_z z)}$ that couples the azimuthal and vertical directions [68-69]. Here, for the first time, we utilize such a similarity for the direct generation of the vortex radiation, which is free from any spatial phase extraction/modulation technology reported before [15, 17, 66-67, 73-78].

The angular phase factor of $e^{il\varphi}$ contained in the expression of cavity field represents a traveling WGM with well-defined OAM of $l\hbar$. However, a single emitter in a normal ring cavity will excite an equal amount of CW (left handed) and CCW (right handed) traveling waves as we have mentioned before, which carry OAM with opposite signs and result in a beam with zero net OAM.

This obstacle is eliminated by using chiral-reversing dipole radiation in a CPN. In the vertical direction $z$, the excited chiral cavity mode is a standing wave consisting both negative and positive $z$-momentum. As the CPN is half encapsulated, the outgoing wave in the positive $z$ direction leads to a vortex radiation to free space as shown in Fig. 3a. As a comparison, Fig. 3b shows the electrical profile of the CCW eigenstate of the CPN, where both the intra-cavity and radiation field have the opposite handedness. This contrast shows clearly the chiral-reversing dipole radiation phenomena. The detailed description of the full wave simulation is given in method and Supplementary Note 3.

**Quantum vortex emitters operated at telecommunication wavelengths**

Based on the principle discussed above, we first designed two quantum vortex emitters with distinct material systems and operation wavelengths and then verified them via full wave simulations. The first one was designed to operate at 1550 nm, where the material system of an InAs quantum dot embedded in InP was adopted [79-81]. The second one was designed to operate at 900 nm, where the material system of InAs quantum dot embedded in GaAs was chosen [81-83]. In the following, we show the result of the first design as an example, while the other one is presented in Supplementary Fig. 5 and 6.

In the design, the height of the CPN is 210 nm and the width of the insulator ring is 50 nm. The inner diameter of the insulator ring is varied by tens of nanometer for the desired OAM in the vortex radiation. The dipole is positioned at $\varphi_0 = \frac{\pi}{2l}$ as required by the condition $\chi_{ba} = -\gamma_{tot}$, and the refractive index modulation is set to $\delta n_I = \delta n_R = 0.003$ (Supplementary Note 3 and Supplementary Fig. 7). We note that the dimensions of the CPN ensure that only one fundamental symmetric plasmonic mode is supported, which is necessarily to achieve a near unity spontaneous emission coupling ($\beta$) factor. The strongly confined electromagnetic field inside the CPN at resonance lead to a high Purcell factor ($F_p$), which will be discussed in detail below (Supplementary Fig. 8). We note that for any practical quantum optics application of a single emitter, $\beta \mapsto 1$ and $F_p \gg 1$ are necessary for both high collection efficiency and the suppression of non-radiative emissions.

Figure 4b shows the simulated far field pattern of the dipole radiation, where most energy is emanated to free space from the upper facet of the cavity. To show its vortex nature, in Figs. 4c and 4d we plot $E_\rho$ and $|\mathbf{E}|$ of the radiation field inside the cavity. $E_\rho$ is the dominant field here and it displays features of a radially polarized WGM with $l = -2$ (Fig. 4c); the uniform $|\mathbf{E}|$ field in the azimuthal direction and the circulating Poynting vector shown in Fig. 4d also indicate that the excited field is indeed a traveling WGM. Furthermore, we plot $E_\rho$ and $|\mathbf{E}|$ at a height of 1550 nm above the cavity in Figs. 4f and 4g. The spiral pattern of $E_\rho$ reveals a phase factor of $e^{i2\varphi}$ (Fig. 4e), and the undefined phase at the center indicates a topological phase singularity on the beam axis (Fig. 4b, Fig. 4e and Supplementary Fig. 9). In addition, the Poynting vector of the emission beam shares the same circulating feature as the field inside the cavity. These results unambiguously confirm that the CPN twists the dipole emission into a vortex beam with a topological charge of $-2$. By tuning the azimuthal order of the WGM, we confirmed that a dipole can also generate vortex emission with other well defined topological charges ($l=-1$ and $-3$ are shown in Supplementary Figs. 10 and 11).

As we have mentioned before, the coupling between the CW and CCW fields inside the cavity depends on $\varphi_0$ and so does the chirality of the dipole radiation, which can be defined quantitatively as [7]

$$\alpha = 1 - \frac{\min[\beta_{\text{CW}}, \beta_{\text{CCW}}]}{\max[\beta_{\text{CW}}, \beta_{\text{CCW}}]}. \tag{8}$$

Here $\beta_{\text{CW(CCW)}}$ is the $\beta$ factor of CW (CCW) field (Supplementary Material Note 4 and Supplementary Fig. 12 and Fig. 13). Figure 5a shows the simulated $\beta$ factors of the dipole radiation at resonance as a function of its position $\varphi_0$. The total $\beta$ factor of these two $|l|=2$ OAM fields ($\beta_{\text{CW}} + \beta_{\text{CCW}}$) approaches unity, as a result of the large Purcell factor and the large free spectrum range of our chiral plasmonic nanocavity. Figure 5b shows the simulated chirality of the dipole radiation, and its maximum is reached at $\varphi_0 = \pi/4$ required by the chiral-reversing condition, where a pure traveling CW wave is excited with $\beta_{\text{CW}} \approx 0.9813$. In Fig. 5d, we show the electric field distribution excited by the dipole at different azimuthal positions. It can be clearly seen that the ratio of the CW and CCW field can be controlled by tuning the azimuthal position of the dipole. The simulation results are in very good agreement with a simple calculation from the coupled mode theory (Solid line in Figs. 4a and 4b).

The spontaneous emission rate ($\gamma$, $\gamma = 1/\tau$, $\tau$: emission lifetime) can be increased by spatial and spectral confinement of the optical field, known as the Purcell effect [84]. A high emission rate is crucial for a quantum dot emitter with large quantum efficiency and emission rate, and it also suppresses the blinking of the quantum dot. The Purcell enhancement factor ($F_P$) is proportional to $Q/V_{\text{mode}}$, where $Q$, $V_{\text{mode}}$ is the quality factor and mode volume of the cavity. Our CPN has an extremely small $V_{\text{mode}}$ of $0.24 \times \left(\frac{\lambda}{2n_{\text{eff}}}\right)^3$ and a mediate $Q$ of 480 (See method). Here we calculate the radiative decay rates acceleration by the cavity which is defined as $\gamma_{\text{emission}}/\gamma_0$, where $\gamma_{\text{emission}}$ and $\gamma_0$ is the radiative decay rate of the dipole in the nanocavity and free space, respectively. Fig. 5c shows $\gamma_{\text{emission}}/\gamma_0$ at varied wavelength under the condition that $\varphi_0 = \pi/4$. At zero detuning, the radiative decay rate of a dipole emission is accelerated by as high as 965 times comparing to the free space radiation. We also calculate the $\gamma_{\text{emission}}/\gamma_0$ by the coupled mode theory (red solid line), which matches well the simulation result.

**DISCUSSION**

In summary, we have revealed the surprising phenomena of chiral-reversing dipole radiation induced by eigenstates phase locking. It takes place when a single dipole emitter is placed in a PT-symmetric ring cavity at its exceptional point, and the singularity of the exceptional point results in a π phase difference between the two coalesced eigenstates when they interact with the dipole emitter. This eigenstates phase locking forces the dipole emitter to radiate into the opposite handedness of the eigenstate, which breaks the conventional wisdom that an emitter will only interact and radiate to eigenstates of the photonic environment. Not only does the chiral-reversing dipole radiation phenomenon enrich the interesting physics of the exception point in single emitter region; it also provides a novel tool to manipulate the single emitter radiation in an unprecedented manner. We further employ such an intriguing phenomena to construct vortex radiation with a controllable topological charge from a single quantum dot emitter with Purcell enhancement factor up to 1000. Our scheme may open a new paradigm for chiral quantum optics and vortex lasers at nanoscale.

**METHODS**

**Full wave numerical simulations.** The simulations are calculated by the finite element electromagnetic solver (COMSOL Multiphysics 5.0, RF module) with tetragonal meshing and scattering boundary conditions.. In 2D simulations, the maximum and minimum element size of different regions is 15/*n* nm and 0.15/*n* nm respectively, where *n* is the real part of the refractive index in different regions. The maximum element growth rate is 1.1, the curvature factor is 0.2, and the resolution of narrow regions is 1. We used the direct MUMPS with a convergence relative tolerance of $10^{-6}$. In 3D simulations, the general maximum element size is 144∕n nm, and the general minimum element size is 1.44∕n nm; the maximum element size in the InP/InGaAsP region is 7.7 nm, and the maximum element growth rate is 1.3. The curvature factor and resolution of narrow regions are the same as the 2D simulations. We used the direct MUMPS with a convergence relative tolerance of $10^{-3}$. In the dipole excited field simulations, since the linewidth of single emitters (~0.01 nm) can be much narrower than the linewidth of the cavity mode, we set the linewidth of dipole source as a delta function. In the simulation, we considered the condition that the temperature is set to be 4.5 K to reduce the metal loss. The refractive index of the material is set as follows: $n_{\text{InP}} = 3.0806$, $n_{\text{cr}} = 3.6683 - 4.18i$, $n_{\text{Ge}} = 4.275 - 0.00567i$, and $n_{\text{Ag}} = 0.0014 + 10.9741i$.

**Numerical calculations of Q values and mode volumes.** The $Q$ value is calculated from the formula $Q = f_r/\Delta f$, where the $f_r$ is the resonance frequency and $\Delta f$ is the full width at half maximum of the resonance spectrum. The mode volume is calculated from $V_m = \frac{W_{\text{total}}}{\max[W(\mathbf{r})]}$, where $W_{\text{total}}$ is the total mode energy integrated over the entire space, i.e., $W_{\text{total}} = \iiint W(\mathbf{r}) d^3\mathbf{r}$. $W(\mathbf{r})$ is the local energy density $W(\mathbf{r}) = \frac{1}{2}(\text{Re}\left[\frac{d(\omega\varepsilon)}{d\omega}\right] |\mathbf{E}(\mathbf{r})|^2 + \mu|\mathbf{H}(\mathbf{r})|^2)$. The peak energy density $\max[W(\mathbf{r})]$ is found by comparing all the energy density in the entire simulation regions. Here, $\varepsilon$ and $\mu$ are permittivity and permeability of the materials, respectively. The dispersion item $\omega\frac{d\varepsilon}{d\omega}$ of Ag is 284.1.

single solid-state spin and a photon, *Nat. Nanotech.* **11**, 539-544 (2016).

84. Purcell, E. M. Spontaneous emission probabilities at radio frequencies, *Phys. Rev.* **69,** 681-681 (1946).

**Acknowledgement**

This work was supported by the National Natural Science Foundation of China (Nos. 11574012, 61521004), the "Youth 1000 Talent Plan", and the National Science Foundation of the United States (No. DMR-1506987).

**Author contributions:**

R.-M.M. developed the concept and supervised the work. X.-Y.W., H.-Z.C., S.W., S.Z. and R.-M.M. designed and performed the mode coupling analysis and numerical simulations. L.G. performed the Green function analysis. R.-M.M., X-Y.W. and H.-Z.C wrote the manuscript. All authors discussed the results and revised the manuscript.

**Competing interests:** The authors declare no competing financial interests.

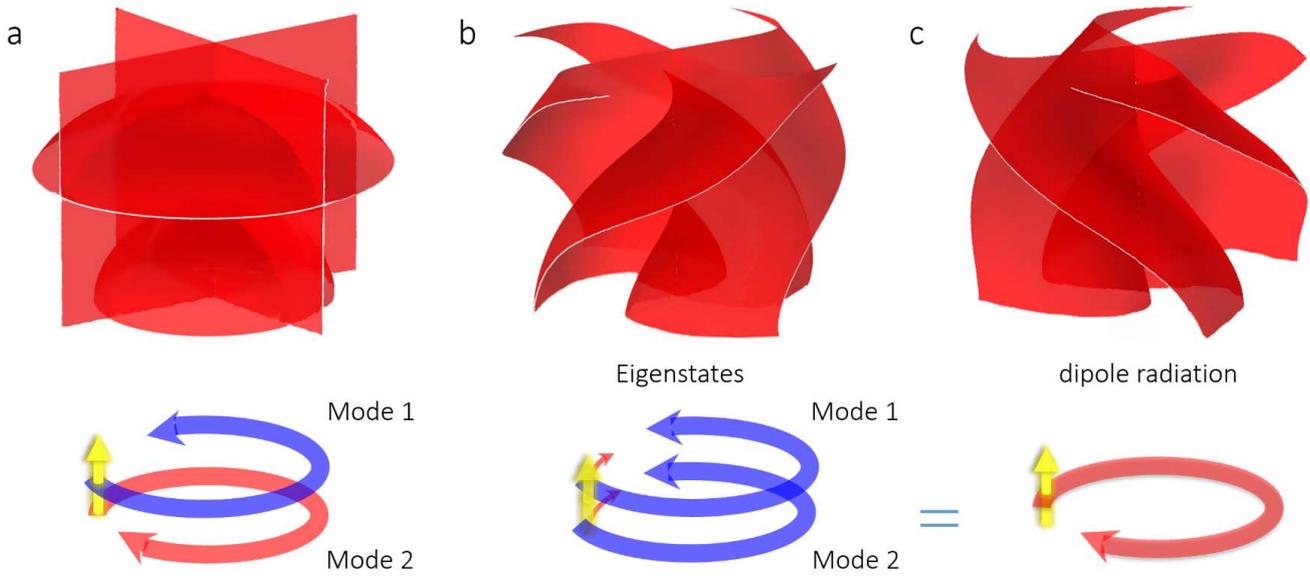

**Figure 1 | Chiral-reversing dipole radiation by eigenstates phase locking. (a)** A dipole emitter inside a normal ring cavity will excite both of the two counter-propagating eigenstates, which forms a standing wave inside the cavity with zero orbital angular momentum. Upper panel: the radiation field pattern (isosurface of $E_r = 0$) of a dipole emitter inside a normal ring cavity. **(b)** A ring cavity operating close to an exceptional point. The two eigenstates become coalesced. Upper panel: the radiation field pattern (isosurface of $E_\rho = 0$) of the coalesced eigenstates. A dipole emitter (yellow arrow) will not radiate to the coalesced eigenstates as intuitive thinking. **(c)** A dipole emitter inside a ring cavity operating close to an exceptional point. The singularity of exceptional point results in a π phase locking of the two coalesced eigenstates when they interact with the dipole emitter (yellow arrow), which forces the dipole emitter to radiate with the opposite handedness of the CCW eigenmode. Upper panel: the radiation field pattern (isosurface of $E_r = 0$) of a dipole emitter inside a ring cavity operating close to an exceptional point.

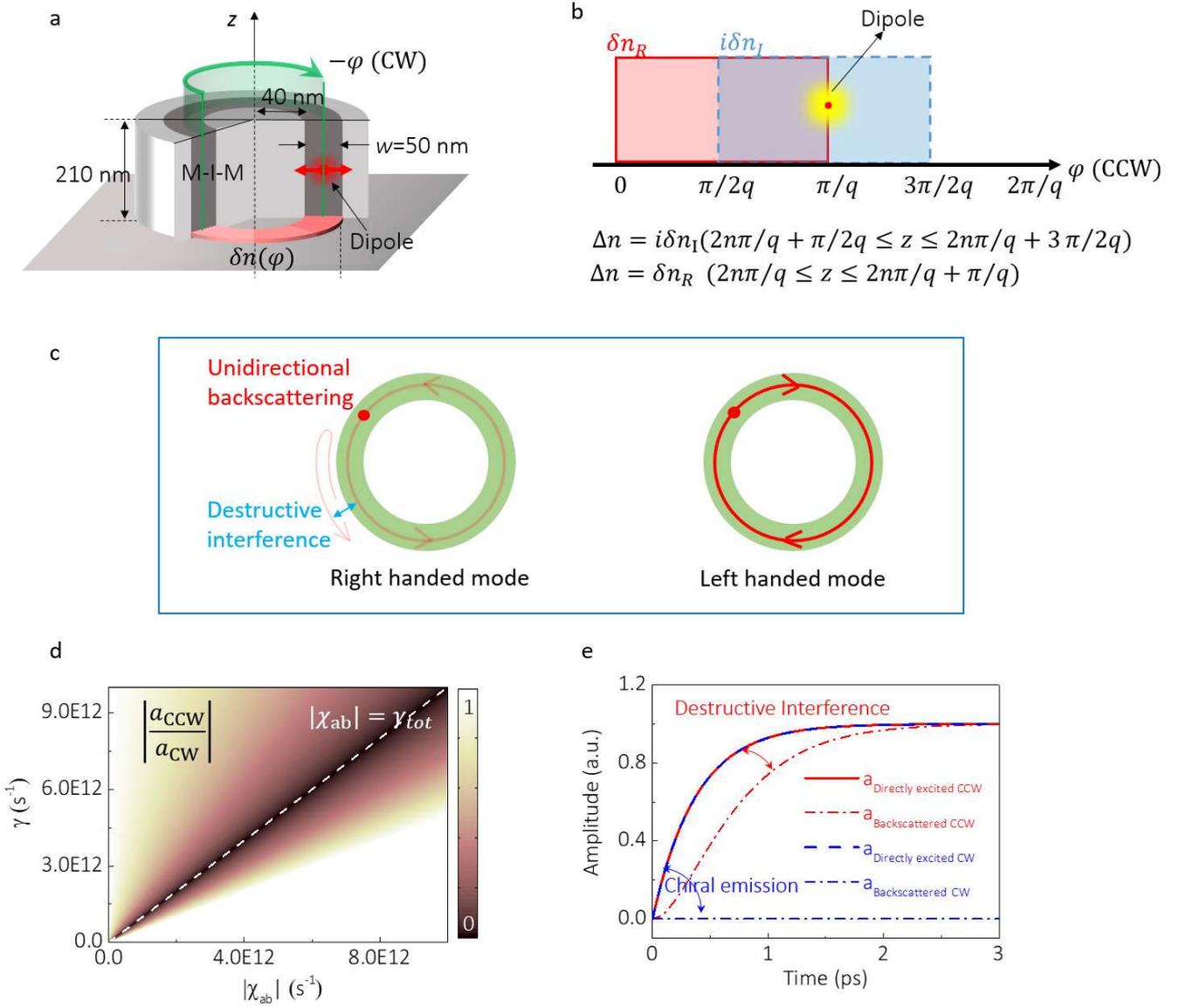

**Figure 2 | Chiral-reversing dipole radiation from a single emitter inside a chiral plasmonic nanocavity.** **(a)** Schematic of the chiral plasmonic nanocavity. M-I-M represents metal-insulator-metal. Cavity size is chosen for the fundamental chiral cavity mode of the material (Supplementary Fig. 10). **(b)** PT symmetric modulation $\delta n(\varphi)$ in chiral plasmonic nanocavity. $\delta n(\varphi)$ is divided into $2l$ periods for WGMs with orbital angular momentum $l$. Each period of $\delta n(\varphi)$ consists of four angularly equidistant parts of $\delta n_R$, $\delta n_R + \delta n_I i$, $\delta n_I i$ and 0 arranging in counterclockwise direction, where $\delta n_R$ and $\delta n_I$ denote the real part and imaginary part of $\delta n(\varphi)$ respectively. Here $q = 2k_0$. **(c)** A dipole inside the chiral plasmonic nanocavity operating at exactly the exceptional point. The chiral-reversing dipole radiation can be

understood from the completely destructive interference between the directly radiated CCW wave and the backscattered CCW wave. **(d)** Amplitude ratio between the CCW and CW waves in the radiation field. **(e)** Time-resolved amplitudes of all the fields the dipole excited inside the chiral plasmonic nanocavity at $|\chi_{ba}|=\gamma_{tot}$. Directly exited and backscattered CCW waves are $\pi$ out of phase and cancel each other in the steady state.

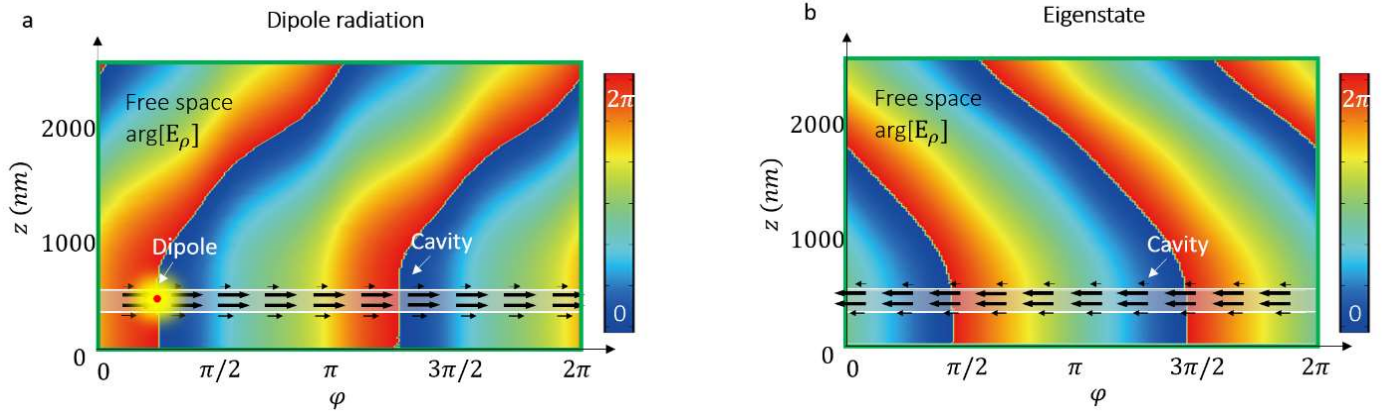

**Figure 3 | Chiral-reversing vortex radiation from a single emitter. (a-b)** Phase distribution of the radial electric field $E_\rho$ of the dipole excited field (a) and the CCW eigenmode (b) in the $z - \varphi$ plane. The black arrows denote the Poynting vector. Both the intra-cavity and radiation field are with opposite handedness between the dipole exited field and the eigenstate, showing clearly the chiral-reversing dipole radiation phenomena.

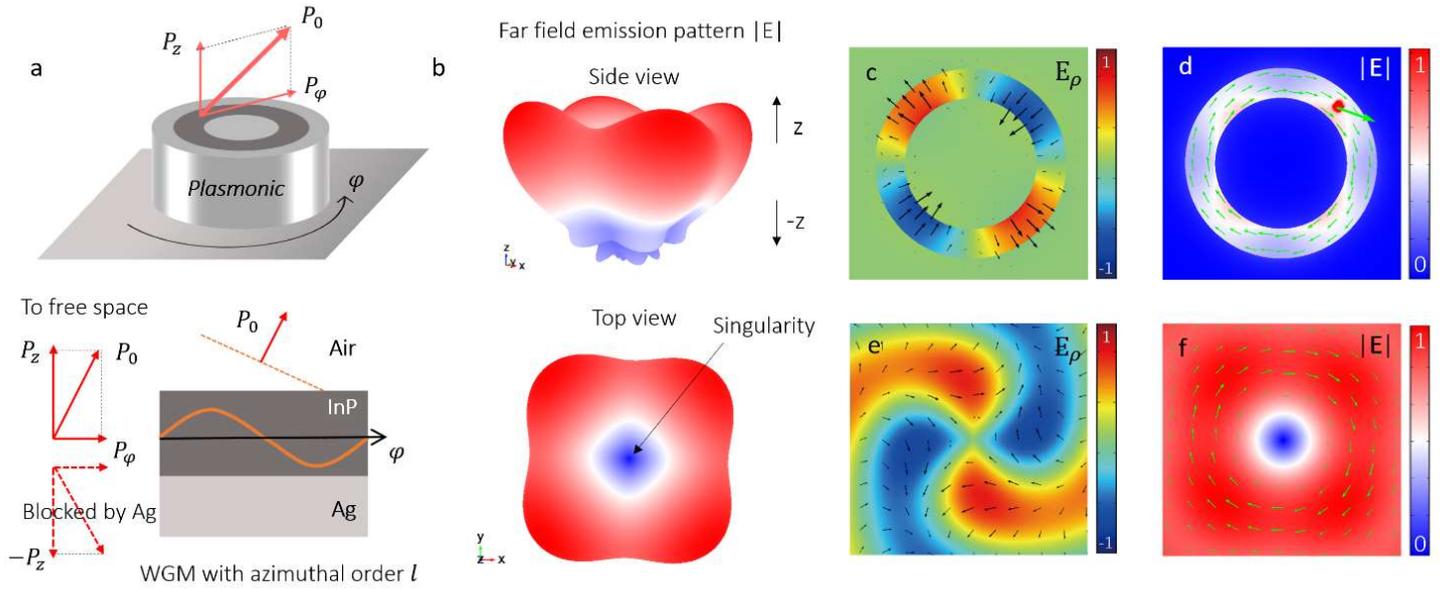

**Figure 4 | Full wave simulations for a quantum vortex emitter operating at telecommunication wavelength**. **(a)** The device is designed to operate at 1550 nm, where an InAs quantum dot is embedded in the middle InP ring (dark) region. Chiral cavity field will couple to free space vortex beam from the open facet of the nanocavity. $P_0$, $P_z$, and $P_\varphi$ represent the total momentum and its $z$ and azimuthal components, respectively. **(b)** Simulated side view (upper panel) and top view (bottom panel) of far field radiation pattern of the dipole excited field. $E_\rho$ and $|\mathbf{E}|$ of the single dipole excited field are plotted inside the cavity (c,d) and at a height of 1550 nm above the cavity (e,f). In (c) - (e), the black and green arrows denote polarization and Poynting vector, respectively. In (f), the green arrows denote azimuthal component of Poynting vector.

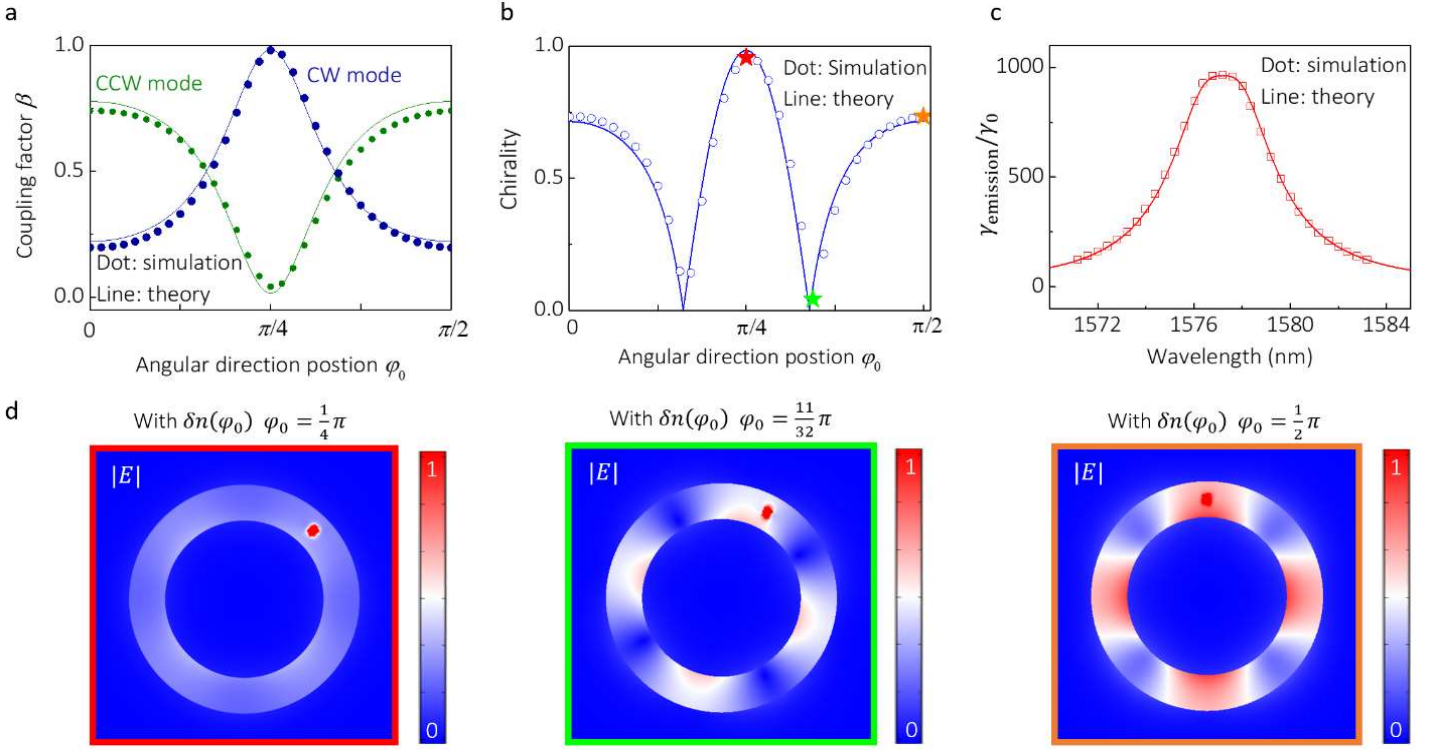

**Figure 5 | Chirality and radiation rate enhancement of a single emitter in the chiral plasmonic nanocavity. (a)** $\beta_{CW}$ and $\beta_{CCW}$ as a function of the dipole position $\varphi_0$ at resonance. **(b)** Chirality of the single emitter radiation at resonance as a function of its position $\varphi_0$. **(c)** Radiation rate enhancement $\gamma_{\text{emission}}/\gamma_0$ at varied wavelength under the condition that $\varphi_0 = \pi/4$. In (a-c), dots and solid line are obtained from full wave simulation and coupled mode theory respectively. **(d)** The electric field excited by dipoles at different azimuthal positions in a cavity with PT symmetric modulation.